\documentclass{article}
\begin{document}
\def\p {{\partial}}
\def\n {{\nu}}
\def\m {{\mu}}
\def\a {{\alpha}}
\def\bt {{\beta}}
\def\f {{\phi}}
\def\th {{\theta}}
\def\g {{\gamma}}
\def\eps {{\epsilon}}
\def\e {{\psi}}
\def\k {{\chi}}
\def\la {{\lambda}}
\def\na {{\nabla}}
\def\bn {\begin{eqnarray}}
\def\en {\end{eqnarray}}
\title{Canonical formulation treatment of a free relativistic
spinning particle}
\maketitle
\begin{center}
\author{S.I. MUSLIH\\Dept. of Physics\\ Al-Azhar University\\
Gaza, Palestine}
\end{center}

\begin{abstract}
The Hamilton - Jacobi method of constrained systems is discussed.
The equations of motion for a free relativistic spinning particle
are obtained without using any gauge fixing conditions. The
quantization of this model is discussed.
\end{abstract}
\section{Introduction}

The theory of constrained systems was developed by Dirac [1,2]
and is becoming the fundamental tool for the study of classical
systems of particles and fields [3,4]. In his method Dirac
distinguish between two types of constraints, first and second
class constraints. As there is an even number of second- class
constraints,these are used to eliminate a certain number of
canonically conjugate pairs of variables. The Dirac Hamiltonian
is then the canonical Hamiltonian plus first class constraints,
which are considered as generators of gauge transformations. To
eliminate the gauge freedom, one has to impose an external gauge
fixing conditions for each first class constraint and evaluate
the new Dirac bracket.

Recently, the canonical method [5-7] has been developed to
investigate constrained systems. The equations of motion are
obtained as total differential equations in many variables which
require the investigation of integrability conditions. If the
system is integrable, one can solve the equations of motion
without using any gauge fixing conditions. In this paper, we
shall treat the free relativistic spinning particle as a
constrained system. In fact, this work is a continuation of
previous paper [5] in which we have obtained the equations of
motion for a free relativistic spineless particle and the
canonical phase space coordinates $q_i$, $p_i$ are obtained in
terms of parameter $x_0$.

Now we would like to give a brief discussion of the canonical
method [5-7]. This method gives the set of Hamilton - Jacobi
partial differential equations [HJPDE] as

\bn
&&H^{'}_{\a}(t_{\bt}, q_a, \frac{\p S}{\p q_a},\frac{\p S}{\p
t_a}) =0,\nonumber\\&&\a, \bt=0,n-r+1,...,n, a=1,...,n-r,\en where
\begin{equation}
H^{'}_{\a}=H_{\a}(t_{\bt}, q_a, p_a) + p_{\a},
\end{equation}
and $H_{0}$ is defined as
\bn
 &&H_{0}= p_{a}w_{a}+ p_{\m} \dot{q_{a}}|_{p_{\n}=-H_{\n}}-
L(t, q_i, \dot{q_{\n}},
\dot{q_{a}}=w_a),\nonumber\\&&\m,~\n=n-r+1,...,n. \en

The equations of motion are obtained as total differential
equations in many variables as follows:

\bn
 dq_a=&&\frac{\p H^{'}_{\a}}{\p p_a}dt_{\a};\\
 dp_a=&& -\frac{\p H^{'}_{\a}}{\p q_a}dt_{\a};\\
dp_{\bt}=&& -\frac{\p H^{'}_{\a}}{\p t_{\bt}}dt_{\a};\\
 dZ=&&(-H_{\a}+ p_a \frac{\p
H^{'}_{\a}}{\p p_a})dt_{\a};\\
&&\a, \bt=0,n-r+1,...,n, a=1,...,n-r\nonumber \en where
$Z=S(t_{\a};q_a)$. The set of equations (4-6) is integrable [5]
if

\bn
dH^{'}_{0}=&&0,\\
dH^{'}_{\m}=&&0,  \m=n-p+1,...,n. \en If conditions (8)and (9) are
not satisfied identically, one considers them as new constraints
and again testes the consistency conditions. Hence, the canonical
formulation leads to obtain the set of canonical phase space
coordinates $q_a$ and $p_a$ as functions of $t_{\a}$, besides the
canonical action integral is obtained in terms of the canonical
coordinates.The Hamiltonians $H^{'}_{\a}$ are considered as the
infinitesimal generators of canonical transformations given by
parameters$t_{\a}$ respectively.

For the quantization of constrained systems we can use the
Dirac's method of quantization [1]. In this case we have
\begin{equation}
H^{'}_{\a}\Psi=0,\;\;\;\a=0,n-r+1,...,n,
\end{equation}
where $\Psi$ is the wave function. The consistency conditions are
\begin{equation}
[H'_{\m}, H'_{\n}]\Psi=0,\;\;\;\m,\n=1,...,r,
\end{equation}
where$[,]$ is the commutator. The constraints $H'_{\a}$ are
called first- class constraints if they satisfy
\begin{equation}
[H'_{\m}, H'_{\n}]=C_{\m\n}^{\g}H'_{\g}.
\end{equation}

In the case when the Hamiltonians $H'_{\m}$ satisfy
\begin{equation}
[H'_{\m}, H'_{\n}]=C_{\m\n},
\end{equation}
with $C_{\m\n}$ do not depend on $q_{i}$ and $p_{i}$, then from
(11) there arise naturally Dirac' brackets and the canonical
quantization will be performed taking Dirac's brackets into
commutators.

On the other hand, The path integral quantization is an
alternative method to perform the quantization of constrained
systems. If the system is integrable then one can solve equations
(4-6) to obtain the canonical phase-space coordinates as
\begin{equation}
q_{a}\equiv q_{a}(t, t_{\m}),\;\;\;p_{a}\equiv p_{a}(t,
t_{\m}),\;\;\m=1,...,r,
\end{equation}
then we can perform the path integral quantization using Muslih
method [8-12] with the action given by (7).

 After this introduction we shall treat the relativistic
spinning particle as a constrained  system and demonstrate the
fact that gauge fixing is solved naturally if the canonical
method is used.

\section{The free relativistic spin particle as a constrained system}

Let us consider the action of the free relativistic spinning
particle as [4]

\bn S =&& \int L {d\tau},\\
L =&& - \frac{{\dot x^{2}}}{2e} + \frac{i {\dot x}\k \e}{e}-
\frac{e m^{2}}{2}- i\e {\dot \e} + i {\e}_{5} {\dot {\e}_{5}} +i
m\k \e_{5}, \en where $x^{\m}$, $e$ are even variables and
$\e^{\m}$, $\k$, $\e_{5}$ are odd variables. With the action (15)
there exist two types of gauge transformations:

\bn \delta\k =&& -\dot{\eps},\;\delta\e_{5} = [
\frac{i}{me}\e_{5}({\dot \e_{5}} - \frac{m}{2}\k)
-\frac{m}{2}]\eps,\nonumber\\
\delta x_{\m} = && i\e_{\m} \eps,\; \delta e = -i\k \eps,\;\delta
\e_{\m}= \frac{1}{2e}(\dot x_{\m}- i \k \e_{\m})\eps, \en
and
\begin{equation}
\delta x= {\dot x}\zeta,\; \delta e = \frac{d}{d \tau}(e\zeta),\;
\delta \e_{\m}={\dot \e_{\m}}\zeta,\; \delta\k=
\frac{d}{d\tau}(\k\zeta),\;\delta \e_{5}= {\dot \e_{5}}\zeta,
\end{equation}
where $\zeta(\tau)$ are even parameters, while $\eps(\tau)$ are
odd parameters. The canonical momenta are defined as

\bn p_{\m}=&& \frac{\p L}{\p {\dot x^{\m}}}=-\frac{1}{e}(\dot
x_{\m}-i \k \e_{\m}),\\
\pi_{e}=&& \frac{\p L}{\p {\dot e}}= 0= -H_{1},\\
\pi_{\k}=&& \frac{\p_{r} L}{\p {\dot \k}}= 0= -H_{2},\\
\pi_{\m}=&& \frac{\p_{r} L}{\p {\dot \e^{\m}}}=-i \e_{\m}=- H_{3},\\
\pi_{5}=&& \frac{\p_{r} L}{\p {\dot \e^{5}}}=i \e_{5}=- H_{4}, \en
where $\p_{r}$ means right derivatives and the metric convention
$g = (+1, -1, -1, -1)$. Now equation (19) leads us to obtain the
velocities $ \dot x_{\m}$ in terms of momenta and coordinates as
\begin{equation}
\dot x_{\m}=(- e p_{\m} + i\k \e_{\m})=w_{\m}.
\end{equation}
The canonical Hamiltonian $H_{0}$ can be obtained as

\bn H_{0}=&&p_{\m}w^{\m}- {\dot e}H_{1} - {\dot \k}H_{2}- {\dot
\e_{\m}}H_{3}- {\dot \e_{5}}H_{4} - L,\\
H_{0}=&& -\frac{e}{2}(p^2 - m^2) +i \k(\e \cdot p -m \e_{5}). \en
Making use of equations (1,2) and (26   ), the set of Hamilton-
Jacobi partial differential equations reads

\bn H'_{0}=&& p^{(\tau)} + H_{0}=0;\;\;\;\; p^{(\tau)}=\frac{\p
S}{\p \tau},\\
H'_{1}=&& \pi_{e}= 0;\;\;\;\;\;\;\;\;\;\;\;\;\;\; \pi_{e}=\frac{\p S}{\p e},\\
H'_{2}=&& \pi_{\k}= 0;\;\;\;\;\;\;\;\;\;\;\;\;\;\; \pi_{\k}=\frac{\p S}{\p \k},\\
H'_{3}=&& \pi_{\m} + i \e_{\m}= 0;\;\;\;\;\;\; \pi_{\m}=\frac{\p S}{\p \e_{\m}},\\
H'_{4}=&& \pi_{5} - i \e_{5}= 0;\;\;\;\;\;\; \pi_{5}=\frac{\p
S}{\p \e_{5}}. \en
This set leads to the total differential
equations as

\bn dx_{\m}=&& (-e p_{\m} + i\k \e_{\m})d\tau,\\
dp_{\m}=&&0,\\
d\pi_{e}=&& -\frac{1}{2}(p^2 - m^2)d\tau,\\
d\pi_{\k}=&& i(\e \cdot p -m \e_{5})d\tau,\\
d\pi_{\m}=&& -i\k p_{\m} d\tau + i d\e_{\m},\\
d\pi_{5}=&& i\k d\tau - i d\e_{5},\\
dp^{(\tau)}=&&0.
\en

To check whether the set of equations (32-38) is integrable or not
let us consider the total variations of $ H'_{0}, H'_{1}, H'_{2},
H'_{3}$ and $H'_{4}$. In fact, the total variations of $H'_{3}$
and $H'_{4}$ determine $d\e_{\m}$ and $d\e_{5}$ in terms of
$d\tau$ as

\bn d\e_{\m}=&& \frac{1}{2} \k p_{\m} d\tau,\\
d\e_{5}=&& \frac{m}{2} \k d\tau. \en
The total variation of
$H'_{1}$ leads to the constraint $H'_{5}$ as
\begin{equation}
H'_{5}= (p^2 - m^2),
\end{equation}
the total variation of $H'_{5}$ is identically zero and the total
variation of $ H'_{2}$ leads to the constraint $H'_{6}$ as
\begin{equation}
H'_{6}= (\e \cdot p -m \e_{5}).
\end{equation}
Taking the total variations of $H'_{6}$ one may obtain

\bn dH'_{6}=&& \e_{\m} dp^{\m} +  p^{\m} d\e_{\m} - md\e_{5},\\
dH'_{6}=&& \frac{1}{2}\k(p^2 - m^2)d\tau. \en Calculations shows
that the total variations of $H'_{0}$ is zero. The set of
equations (32-38) is integrable and the phase space coordinates
$(x_{\m}, \e_{\m}, \e_{5})$ and $(p_{\m}, \pi_{\m}, \pi_{5})$ are
obtained in terms of independent parameters $(\tau, e, \k)$.
$H'_{0}, H'_{1}$ and $H'_{2}$ can be interpreted as infinitesimal
generators of canonical transformations given by parameters
$\tau, e$ and $\k$ respectively.

Now we would like to discuss the operator quantization of the
free relativistic spinning particle. The anti commutator
relations corresponding to the pseudo-classical brackets of
$\e_{\m}$ and $\e_{5}$ are
\begin{equation}
[\e_{\m}, \e_{\n}] = g_{\m \n},\;\; [\e_{5}, \e_{5}]= -1,\;\;
[\e_{\m}, \e_{5}]=0,
\end{equation}
which may realized as
\begin{equation}
\e_{\m} =\frac{1}{\sqrt{2}}\g_{5}
\g_{\m},\;\;\e_{5}\frac{1}{\sqrt{2}}\g_{5}.
\end{equation}
Here $ \g_{5}= \g_{0} \g_{1} \g_{2}\g_{3}$, such that
$\g_5^{2}=-1$, $[\g_{5}, \e_{\m}]_{+}=0.$

Consider the physical Hilbert space as a subspace of a Hilbert
space $\Re$ in which the fields $x^{\m}$, $\e_{\m}$, $\e_{5}$ and
their momenta are not submitted to the constraint. Then only
those states of $\Re$ which fulfill

\bn
&&(p^2 - m^2)\mid\e_{phys}\rangle=0,\\
&&\g_{5}(p_{\m} \g^{\m} - m)\mid\e_{phys}\rangle=0, \en belong to
the physical Hilbert space. One should notice that equation (48)
is nothing but the Dirac equation.

Now to obtain the path integral quantization of this system, we
can use equation (7) to obtain the canonical action as \bn
S=&&\int\{[\frac{e}{2}(p^2 - m^2) -i \k(\e \cdot p -m \e_{5}
+p_{\m}(- e p^{\m} + i\k \e^{\m})]d\tau\nonumber\\ +&& [-i\e_{\m}
+\pi_{\m}]d\e_{\m} + [i\e_{5} + \pi_{5}]d\e_{5}\}. \en Making use
of the equations of motion (39,40) and using the definition of
the canonical momenta, one can recover the original action.

\section{ Conclusion}

A free relativistic spinning particle is treated as a constrained
system. The canonical method [5- 7] treatment of this system
leads us to obtain the equations of motion as total differential
equations in many variables. Since the integrability conditions
are satisfied, this system is integrable. Hence, one can solve
the classical dynamics of this system in terms of parameters
$\tau, e, \k $ without using any gauge fixing conditions. Although
 $e$ and $\k$ are treated as coordinates in the Lagrangian, the
 presence of constraints and the integrability conditions force us
 to treat them as parameters like $\tau$.

 The operator quantization of the relativistic spinning particle
 leads to obtain the Dirac equation, which describe simultaneously
 the particle and the anti particle.

 Unlike conventional methods one can  perform the path integral quantization
 of this system using
 Muslih method to obtain the action directly without
 considering any Lagrange multipliers.

 In fact, Dirac's method treatment needs gauge fixing conditions
 to determine the classical dynamics of the constrained system. In
 the the free relativistic spinning particle system, since the
 number of the first class constraints are four, one has to impose
 four supplementary gauge fixing conditions.

\end{document}